\documentclass[12pt,a4paper]{article}
\usepackage{amsmath,amssymb}

\begin{document}

\textwidth=135mm
 \textheight=200mm
\begin{center}
{\bfseries Self-Dual Connections and the Equations of Fundamental Fields in a Weyl-Cartan Space}
\vskip 5mm
V. V. Kassandrov$^{\dag}$~\footnote{E-mail: vkassan@sci.pfu.edu.ru} and J. A. Rizcallah$^\ddag$~\footnote{E-mail: joeriz68@gmail.com}
\vskip 5mm
{\small {\it $^\dag$ Institute of Gravitation and Cosmology, Peoples' Friendship
	University of Russia, 117198, Moscow, Russia}} \\
{\small {\it $^\ddag$ Lebanese University, School of Education, Beirut, Lebanon}}\\

\end{center}
\vskip 5mm
\centerline{\bf Abstract}
\noindent
{\small Spaces with a Weyl-type connection and torsion of a special type induced by the structure of the differentiability conditions in the algebra of complex quaternions are considered. The consistency of these conditions implies the self-duality of curvature.
The Maxwell and  $SL(2,\mathbb C)$ Yang-Mills fields associated with the irreducible components of the connection also turn out to be self-dual, so that the corresponding equations are fulfilled on the solutions of the generating system. Using the twistor structure of
the latter, its general solution is obtained. The singular locus has a string-like (particle-like) structure and generates the self-consistent algebraic dynamics of the string system.}


\vskip 10mm
The most general {\it geometrodynamic approach} to constructing the fundamental dynamics of fields/particles in practice proves to be unproductive due to the absence of clear criteria for selecting the very geometry as well as the arbitrary choice of a Lagrangian. On
the other hand, building the theory on some exceptional algebraic structure, primarily of the quaternion type, we  hope to uniquely determine the spacetime geometry induced by it and  derive the equations of fundamental quaternion fields from the intrinsic
properties of the original algebra alone.

 Such an approach, proposed in \cite{a1,a2} (for more recent work, see \cite{a3}), is based on the differentiability conditions for functions of an algebraic ($\mathbb A$) variable. The above conditions are just a generalization of the Cauchy-Riemann conditions in complex analysis to the case of an associative but noncommutative algebra $\mathbb A$. Such conditions are stated in a component-free form covariant with respect to the multiplication in the algebra $\mathbb A$
 \begin{equation}\label{diff}
dF=\Phi \cdot dZ \cdot \Psi,
\end{equation}
where $F=F(Z)\in \mathbb A$ is an $\mathbb A$-valued function of the variable $Z\in \mathbb A$, $\Phi=\Phi(Z),~\Psi=\Psi(Z)$ are auxiliary $\mathbb A$-valued functions (``{\it semiderivatives}'' of the principal function $F(Z)$), and $dF$ is the linear part of the increment (differential)
of $F(Z)$, corresponding to the increment of the argument $dZ$. 

In previous works, the algebra of complex quaternions (biquaternions)  $\mathbb B$, isomorphic to the full algebra of $2\times2$ matrices over $\mathbb C$, was used as a ``{\it space-time algebra}''  $\mathbb A$. However, the $4\mathbb C$-coordinate space of $\mathbb B$ was reduced to the subspace of Hermitian matrices $Z\mapsto X=X^+$ with Minkowski metric, so that the entire construction turned out to be Lorentz-invariant. The physical fields {\it associated with differentiable $\mathbb B$-functions} remain  $\mathbb C$-valued. Besides, the main class of solutions of (\ref{diff}) corresponds~\cite{a2} to the case of the equality $\Psi(Z)=F(Z)$ (or $\Phi(Z)=F(Z)$); therefore, the  $\mathbb B$-differentiability conditions, which {\it play the role of primary field equations}, finally take the form
\begin{equation}\label{GSEF}
dF = \Phi \cdot dX \cdot F .
\end{equation}

In the nontrivial case of $\mathbb B$-functions (matrices) with ${\det F}=0$ we decompose (\ref{GSEF})  by columns to obtain the so-called {\it generating system of equations} (GSE)
\begin{equation}\label{GSE}
d\xi = \Phi dX \xi
\end{equation}    
for the 2-spinor $\xi=\{\xi_A(X)\}$ and 4-vector $\Phi=\{\Phi_{AA^\prime}(X)\}$ fields. Here $A,B,...A^\prime,B^\prime...=0,1$, and the sign of $\mathbb B$ (matrix) multiplication is omitted. 

It is precisely the system of equations (\ref{GSE}) (purely algebraic) that was used to construct {\it algebrodynamics}, that is, the dynamics of fields and the corresponding singularities, which are treated as particle-like formations. Below we present a number of the most
important properties and consequences of system (\ref{GSE}), referring the reader to the authors' works for proofs (see references in~\cite{a3}).

By eliminating the field  $\Phi(X)$, we reduce the GSE to the system of equations of {\it shear-free null congruences} (SFCs) $\xi^{A^\prime}\partial_{AA^\prime} \xi_C =0$. By analogy with the SFC equations, the {\it general solution} of the original GSE is given by
\begin{equation}\label{Kerr}
\Pi^{(C)}(\xi^{A^\prime},\tau_A)=\Pi^{(C)}(\xi^{A^\prime},X_{AA^\prime} \xi^{A^\prime}) =0,~~~C=1,2
\end{equation}
where $\Pi^{(C)}$ is a pair of arbitrary (holomorphic) functions of the {\it twistor} argument with 4 complex (spinor) components related by the {\it twistor incidence relation} 
$\tau_A = X_{AA^\prime} \xi^{A^\prime}$. At each point $X\in \bf M$, solving system (\ref{Kerr}) for $\xi^{A^\prime}$, we arrive at the 2-spinor field $\xi(X)$ (in general multivalued), each continuous branch of which satisfies the two fundamental relativistic equations: the complex eikonal equation $\partial_{AA^\prime} \xi_C~ \partial^{AA^\prime} \xi_C = 0$ (for each spinor component) and the wave equation $\Box G = 0$ (for the quotient of components $G:=\xi_1 / \xi_0$). Note that (\ref{Kerr}) is an invariant generalization of the so-called
{\it Kerr theorem} for describing SFCs.

As to $\Phi(X)$, it is essentially a {\it gauge} field, since the GSE is form-invariant under transformations of the form 
\begin{equation}\label{gauge}
\xi \mapsto \alpha \xi, ~~\Phi_{AA^\prime} \mapsto \Phi_{AA^\prime} - \partial_{AA^\prime} \alpha, 
\end{equation}
where the gauge parameter $\alpha=\alpha (\xi, X\xi)$ depends on $X$ only through the components of the transformed twistor ${\bf W}=\{\xi, X\xi\}$ (the so-called ``{\it weak gauge invariance}''~\cite{a4}).

Moreover, the compatibility conditions of the overdetermined GSE (\ref{GSE}) $dd\xi = 0 = R\xi, ~R:=(\Phi dX \Phi)\wedge dX$ imply the {\it self-duality} of the curvature 2-form of the effective connection $\Omega:=\Phi dX$ on the GSE solutions (the so-called weak self-duality~\cite{a4}). Even on a flat space-time background, the connection $\Omega$ already possesses nonmetricity (of the Weyl type) and torsion of a specific form~\cite{a4}. The very GSE equations (in form (\ref{GSEF})) can be considered as the equations
of {\it covariantly constant} fields in the corresponding space with the Weyl-Cartan connection~\cite{a2},  $dF=\Omega F$. Note that covariantly constant fields in Weyl-Cartan affine connection spaces of different types can be generally used for promising geometric
interpretations of electromagnetism~\cite{a5}.

Owing to the self-duality of the curvature of the $GL(2,\mathbb C)$ connection $\Omega$ on the GSE solutions, {\it the equations of source-free gauge} fields are fulfilled: Maxwell equations for the {\it trace} of the curvature and the $SL(2,\mathbb C)$ Yang-Mills equations for its traceless part~\cite{a6}. The associated electromagnetic and Yang-Mills fields are singular at the points
\begin{equation}\label{sing}
P:=\det \Vert \frac{d\Pi^{C}}{d\xi^{A^\prime}} \Vert =0,
\end{equation}
which correspond to multiple roots of (\ref{Kerr}) for the spinor $\xi^{A^\prime}$.  If the corresponding singular subsets are bounded in three-dimensional space, they can be considered as particle-like formations, which even possess some properties of quantum particles. For example, the electric charge is a multiple of the minimal (``elementary'') charge, that of a Kerr-Newman singular ring. Solutions with electrically neutral singularities have been obtained too.

From systems (\ref{Kerr}) and (\ref{sing}) it follows that such formations have generally the nature of {\it closed strings}. Thus, the system of algebraic equations (\ref{Kerr}),(\ref{sing}), which follows from the GSE, determines the nontrivial {\it algebraic dynamics of a system of closed strings on} $\bf M$.

In a particular case, such a system degenerates into a locus of singular points on a single worldline, in the spirit of the well-known Wheeler-Feynman concept of the ``{\it one-electron Universe}''. Such collective algebraic dynamics has been studied in detail in~\cite{a7}. At
least in the case of an arbitrary {\it polynomial worldline}, it turns out to be conservative: a set of Lorentz-invariant conservation laws holds. There are also other physically interesting universal properties of the dynamics of the roots of the original algebraic system,
including their (asymptotic) {\it merging and clustering}. The more complicated general case of string algebrodynamics is yet to be studied.

\end{document}